\documentclass[amsmath,amssymb,aps,pre,twocolumn,superscriptaddress]{revtex4-2}
\usepackage{amsmath,amsfonts}
\usepackage{soul}
\usepackage{dsfont}
\usepackage{graphicx}
\usepackage{bm}
\usepackage{hyperref}
\usepackage{url}
\usepackage[english]{babel}
\usepackage[utf8x]{inputenc}
\usepackage[T1]{fontenc}
\usepackage{color}
\usepackage{mhchem}
\usepackage{mhequ}

\begin{document}
\title{Active self-disassembly enhances the yield of self-assembled structures}

\author{Karsten Kruse}
\email{Karsten.Kruse@unige.ch}
\affiliation{Department of Biochemistry, University of Geneva, Geneva, Switzerland}
\affiliation{Department of Theoretical Physics, University of Geneva, Geneva, Switzerland}

\author{Jean-Pierre Eckmann}
\email{Jean-Pierre.Eckmann@unige.ch}
\affiliation{Department of Theoretical Physics, University of Geneva, Geneva, Switzerland}
\affiliation{Section de Mathématiques, University of Geneva, Geneva, Switzerland}

\author{Wilson C.~K.~Poon}
\email{w.poon@ed.ac.uk}
\affiliation{School of Physics \& Astronomy, The University of Edinburgh, Peter Guthrie Tait Road, Edinburgh EH9 3FD, Scotland, United Kingdom}

\date{\today}
\begin{abstract}

We introduce a lattice model to probe the effect of active self-disassembly on equilibrium self-assembly. Surprisingly, we find conditions under which active self-disassembly enhances the yield of a target structure above that achieved by self-assembly alone when the latter is already favoured thermodynamically.  We discuss biological implications of our findings. 
\end{abstract}

\maketitle

Caspar and Klug introduced `self-assembly' in 1962 to describe the synthesis of bacterophage capsids~\cite{Caspar1962,Bruinsma2021}. The idea was picked up in the mid-1980s by the then nascent field of soft matter. Inspired initially by the self-assembly of collodial hard spheres into crystals~\cite{PvM}, soft matter scientists have since then learnt to design moieties such as patchy colloids~\cite{Li2020} that self-assemble under Brownian motion to yield increasingly complex structures~\cite{Whitesides2002}. Meanwhile, functionalised self-assembled hard sphere colloidal crystals have found application as biosensors~\cite{Inan2017}. 

To appreciate the challenge of designer self-assembly, consider synthesising a product with composition $\rm A_mB_n$ and geometry $k$ from components A and B:
\begin{equation}
    \ce{mA + nB <=> (A_{\rm m}B_{\rm n})^{(k)}}. \label{eq:scheme1}
\end{equation}
In simulations and experiments, the final state is typically a mixture of $({\rm A}_{\rm m}{\rm B}_{\rm n})^{(k)}$ and other species $\{({\rm A}_{\rm p}{\rm B}_{\rm q})^{(i)}|{\rm (p,q) \neq (m,n)},i \neq k\}$ that have equal or nearly-equal free energy. Optimising reactant shapes, interactions, etc.~is necessary to maximise the equilibrium constant of the desired reaction relative to all others. This is cumbersome for complex products, because the number of `mistakes' combinatorially multiplies as the free-energy landscape complexifies~\cite{Zeravcic2017}. A bio-inspired solution is to move from equilibrium to non-equilibrium, {\it active} self-assembly, which uses external energy to  drive the desired reaction forward, leaving all other pathways to occur under near-equilibrium conditions~\cite{De2018,Das2021}. 

Significantly, biological self-assembly is invariably coupled to self-{\it dis}assembly processes, which are also often driven off equilibrium by energy dissipation. Thus, in kinetic proofreading, energy is used to improve the accuracy of self-assembly~\cite{Hopfield1974,Ninio1975,Alon2019,Boeger2022}, e.g., in protein translation. This process has been mimicked, e.g, in the {\it in silico} self-assembly of colloids~\cite{Zhu2023}. Separately, `test-tube evolution' experiments involving self-assembled polymers are found to proceed only when active self-disassembly is used to destroy selected small oligomers~\cite{Yang2021}. 

Biological active self-disassembly, e.g.~ATP-dependent proteolysis~\cite{Bittner2016,Feng2021}, is long known and studied. However, except for kinetic proofreading~\cite{Hopfield1974,Ninio1975,Alon2019,Boeger2022}, the {\it logic} of active self-disassembly remains obscure. For example, why cells have evolved energy-dependent proteases alongside ATP-independent protein degradation has seldom been asked, and only tentative, qualitative answers been offered~\cite{Gottesman1992}. The generic principles governing coupled biological active self-disassembly and self-assembly also remain unknown. Progress should impact not only  biology but also materials science, where such coupling may hold the key to next-generation complexity~\cite{Zeravcic2017}.

We study a simple model to probe the logic of active self-disassembly and its interaction with self-assembly. While it is not directly relatable to any physical or biological system, it may uncover generic properties.  Surprisingly, we find conditions under which active self-disassembly enhances the yield of a target structure above that achieved by self-assembly alone when the latter is already favoured thermodynamically.  We discuss biological implications of our findings. 

\begin{figure}
\centering  
\includegraphics[width=0.45\textwidth]{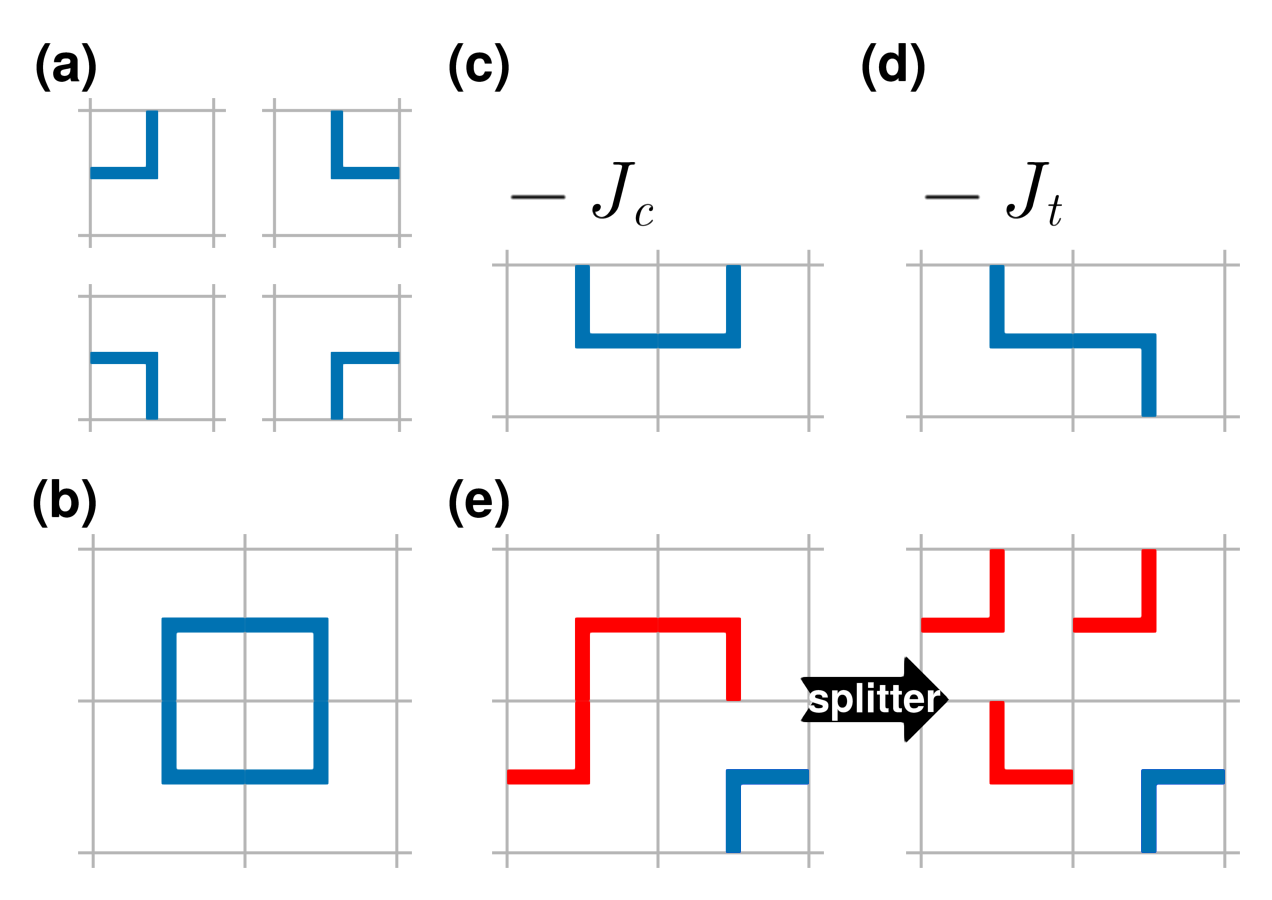}
\caption{Illustration of the model. (a) The four states a site can be in. (b) A loop. (c) Example of a cis-joint: free ends point in the same direction, energy is $-J_c$. (d) Example of a trans-joint: free ends pointing in opposite directions, energy is $-J_t$. (e) Action of a splitter on a chain (red): each site occupied by the chain is assigned a random state.}\label{fig:model}
\end{figure}

We consider an $N\times N$ lattice with periodic boundary conditions. Each site $(i,j)$ can be in one of 4 states (`angles') $\sigma_{i,j}$, Fig.~\ref{fig:model}(a), where $i,j=1,\ldots,N$ enumerate rows and columns with the site $(i+1,j+1)$ being to the top right of the site $(i,j)$. Our target is to self-assemble square loops (hereafter just `loops'), Fig.~\ref{fig:model}(b), by evolving the lattice state $\left\{\sigma_{i,j}\right\}$ in two ways. First, in thermal evolution, two adjacent joined angles have a lower energy than disjoint ones. Specifically, cis- and trans-joins, Fig.~\ref{fig:model}(c, d), have energies $-J_c$ and $-J_t$ respectively, where $J_{c,t} > 0$ are in units of $k_BT$ (with $k_B$  the Boltzmann constant and $T$ the absolute temperature). When $J_c=J_t$, the system's Hamiltonian reduces to that of a vertex model~\cite{Baxter.1982} for spin ice~\cite{Bramwell_2020,Cugliandolo.2017} if we map each join to an incoming leg and each non-join to an outgoing leg. We call an open-ended ensemble of joined sites a `chain'.

Thermal evolution occurs via Glauber dynamics. At each time step, we make a Monte-Carlo (MC) move on a site $(i,j)$ that is chosen uniformly from the lattice. Its state is updated to $\sigma^\prime_{i,j}$, which is accepted with probability $p=\exp\left\{-\Delta E\right\}/\left(1 + \exp\left\{-\Delta E\right\}\right)$, where $\Delta E=E^\prime-E$ is the difference between the energy $E^\prime$ after and $E$ before the update. In this scheme, the system eventually reaches thermodynamic equilibrium.

\begin{figure}
\centering  
\includegraphics[width=0.49\textwidth]{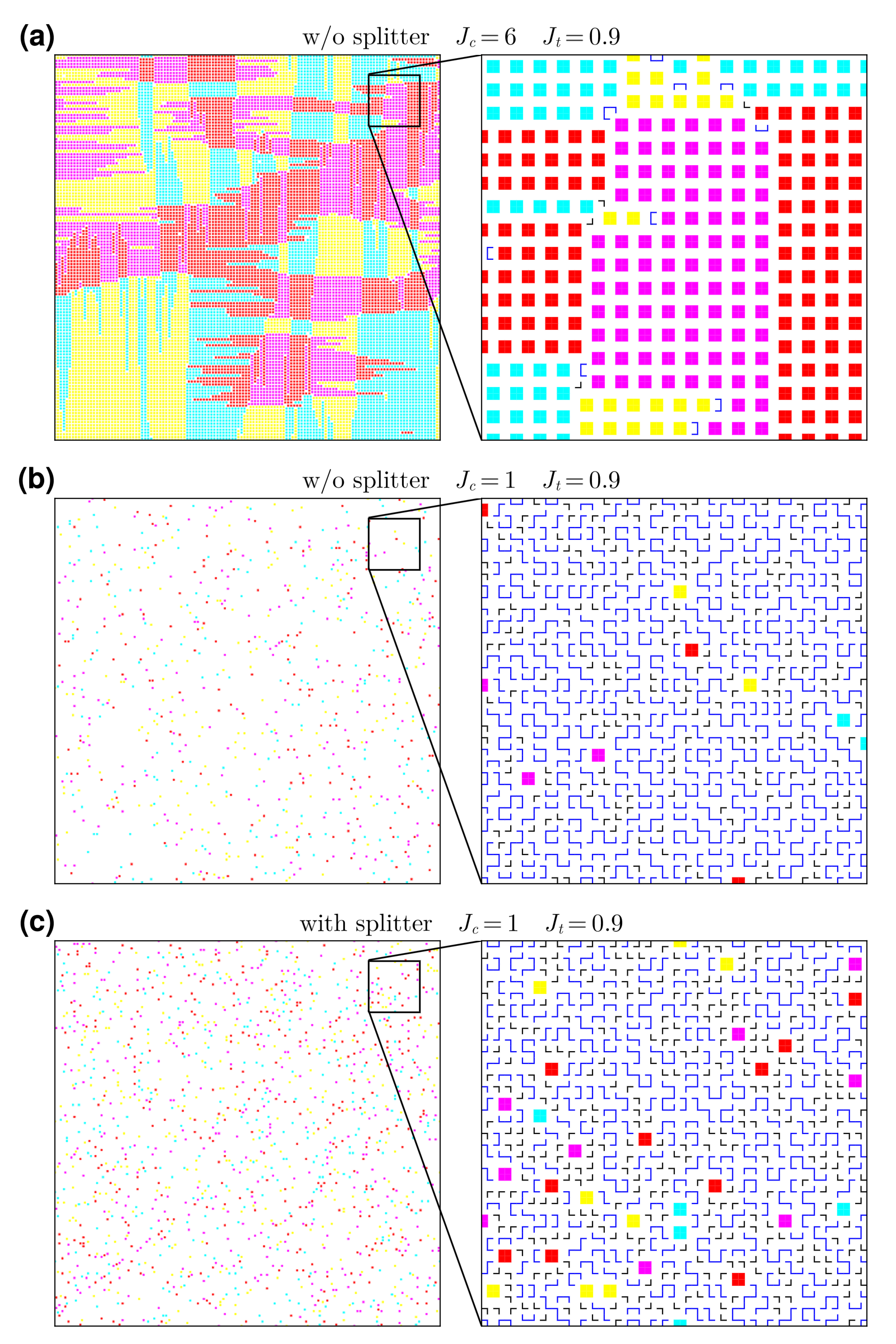}
\caption{Steady-state sample configurations without (a,b) and with splitter (c) for the full $N\times N$ system where $N=256$ (left) and for magnifications (right). Loops on the four different sublattices are shown in red, yellow, magenta, and cyan, isolated angles in black, and chains in blue. For clarity, only loops are shown for the full systems. The samples were taken after approximately $4\cdot10^{10}$ (a) and $10^8$ (b,c) MC steps.}\label {fig:states}
\end{figure}

To produce a non-equilibrium state, we use `splitters' that recognise and disassemble chains but not loops. After each MC step, a splitter `lands' on a random site $(i,j)$. If and only if this site is part of a chain, it is assigned a new state picked from the 4 possibilities with uniform probability. All other sites of the chain are similarly assigned new states, Fig.~\ref{fig:model}(e), so that splitter action is non-local for chains of length $\geq 3$. This active step is athermal and breaks detailed balance, and so will require energy in any physical implementation; but we do not account for this explicitly  in our model. Sample configurations are shown in Fig.~\ref{fig:states}.  

\begin{figure}
\centering
\includegraphics[width=0.45\textwidth]{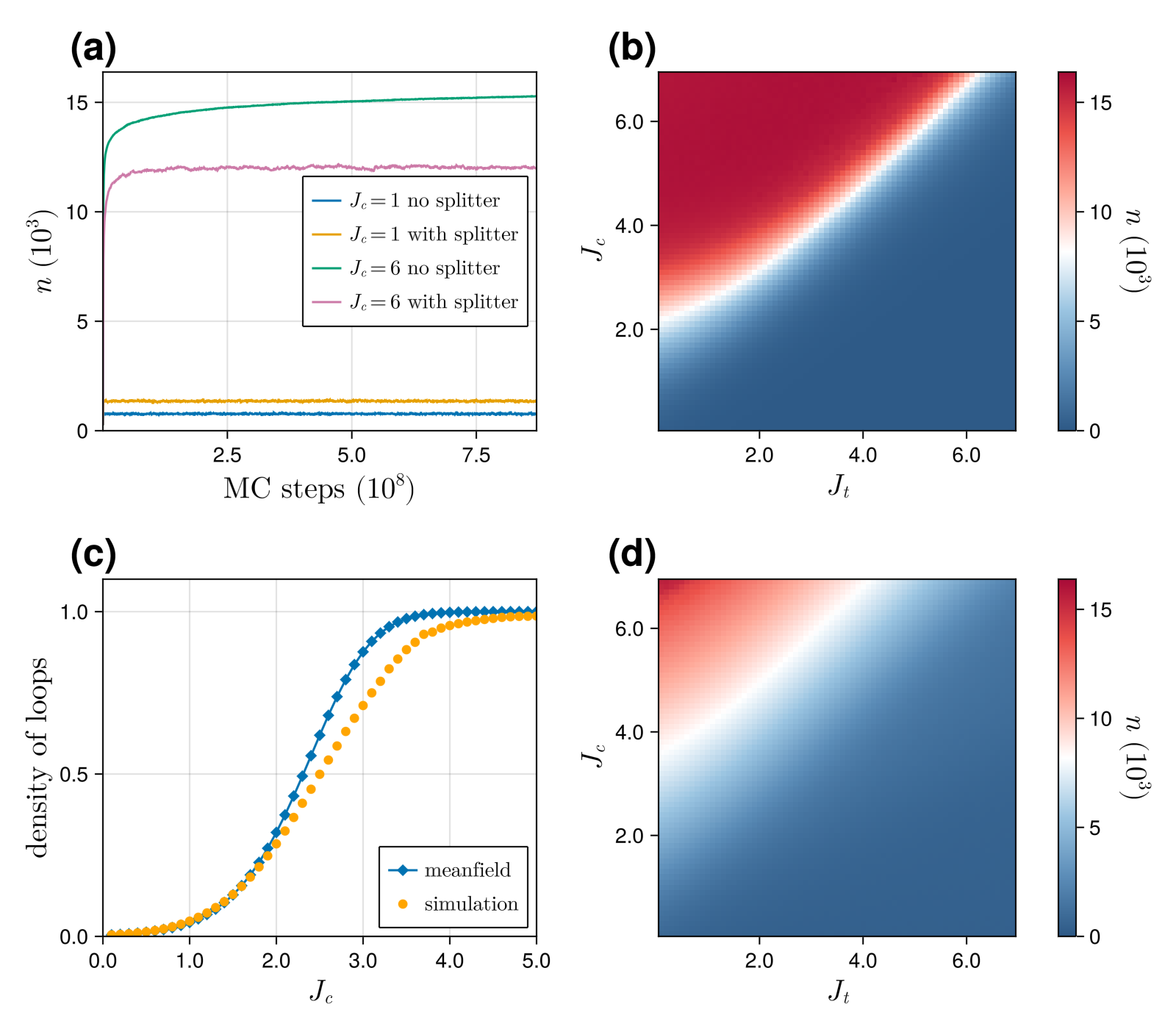}
\caption{Number of loops. (a) Number of loops $n$ as a function of MC steps without and with splitter. (b) Steady state number of loops $n$ without splitter as a function of $J_c$ and $J_t$ from simulations. (c) Steady state density of loops without splitter from simulations and mean-field calculations, see text. (d) Steady state number of loops $n$ with splitter as a function of $J_c$ and $J_t$ from simulations. All simulations were done on a grid with $N=256$, for which the maximum number of loops is $n_\mathrm{max}=16384$. We used $J_t=0.9$ in (a,c).}
\label {fig:parameterDependence}
\end{figure}

Consider first the system without a splitter. It starts from a random initial configuration with each site being in one of the four possible states with probability $\frac{1}{4}$, giving on average $N^2(\frac{1}{4})^4$ loops. In steady state, the average number of loops grows with increasing values of $J_c$, Fig.~\ref{fig:states}(a,b). For join energies on the order of thermal energy, there are only few isolated angles. Similarly, chains with more than 5 joins appear rarely. Loops exist on 4 sublattices. Domains of loops on one sublattice can block the expansion of domains of loops on another sublattice, Fig.~\ref{fig:states}(a), so that equilibration slows with increasing number of loops, Fig.~\ref{fig:parameterDependence}(a). 

As expected, the number of loops increases with increasing $J_c$ and decreasing $J_t$, Fig.~\ref{fig:parameterDependence}(b), because loops, with only cis-joins, are favored by large $J_c$ and low $J_t$. Increasing $J_c/J_t$ saturates the number of loops at $N^2/4$; it is not, however, merely a function of $J_c/J_t$, but depends on the absolute values of both parameters.

A mean-field analysis provides insight into the system's behavior at small loop density. We compute the probability that a $2\times 2$ block of sites is in the loop configuration while assuming that a site adjacent to this block is in any of the four possible states with probability $\frac{1}{4}$. The energy of a configuration $\mathcal{S}=\left\{\sigma_{i,j}\right\}_{i,j=1,2}$ is then
\begin{align}\label{eq:2}
E_\mathcal{S}(J_c,J_t)&=E_{\mathcal{S},\mathrm{in}} + E_{\mathcal{S},\mathrm{out}}.  
\end{align}
Here, without loss of generality, we consider a $2\times 2$ block with the lower left site at $(1,1)$. In Equation~\eqref{eq:2}, $E_{\mathcal{S},\mathrm{in}}$ and $E_{\mathcal{S},\mathrm{out}}$ are the energies due interactions between sites within the block and between sites within and their adjacent sites outside the block, respectively. Explicitly, $E_{\mathcal{S},\mathrm{in}}$ is given by the number of cis-joins in the block times $-J_c$ plus the number of trans-joins in the block times $-J_t$. Each unpaired end of an angle pointing to a site outside the block contributes $-(J_c+J_t)/4$ to $E_{\mathcal{S},\mathrm{out}}$. The probability for a block to be in the loop configuration is then $p=\mathrm{e}^{-E_\mathrm{loop}}/Z$, where $E_\mathrm{loop}=-4J_c$ is the energy of the loop configuration and $Z=\sum_\mathcal{S}\mathrm{e}^{-E_\mathcal{S}}$.

Since the lower left site of the block considered above was fixed at site $(1,1)$, one can interpret $p$ to be the probability of a block on one sublattice forming a loop. Given four sublattices, the total probability to form a loop is $p_l=p+p(1-p)+p(1-p)^2+p(1-p)^3$ with $4n/N^2\to p_l$ for sufficiently large lattices, where $n$ is the number of loops. To obtain this result, consider each sublattice in turn. Then, each term gives the probability that a loop forms on one sublattice times the probability that it has not formed on the sublattices considered before. The mean-field result agrees with simulations when $J_c\lesssim1$, but deviates at larger $J_c$ because loops on difference sublattices interact, Fig.~\ref{fig:parameterDependence}(c).

We next activate a splitter to drive the system out of equilibrium. For small values of $J_c$ and $J_t$, the number of loops is larger than compared to the equilibrium case, Fig.~\ref{fig:states}(c). Furthermore, fluctuations around the steady-state value of $n$ tend to increase, Fig.~\ref{fig:parameterDependence}(a). Qualitatively, the dependence of the steady-state loop number on $J_c$ and $J_t$ is similar in the presence and the absence of a splitter, but differs quantitatively, Fig.~\ref{fig:parameterDependence}(b,d). 

\begin{figure}
\centering\includegraphics[width=0.45\textwidth]{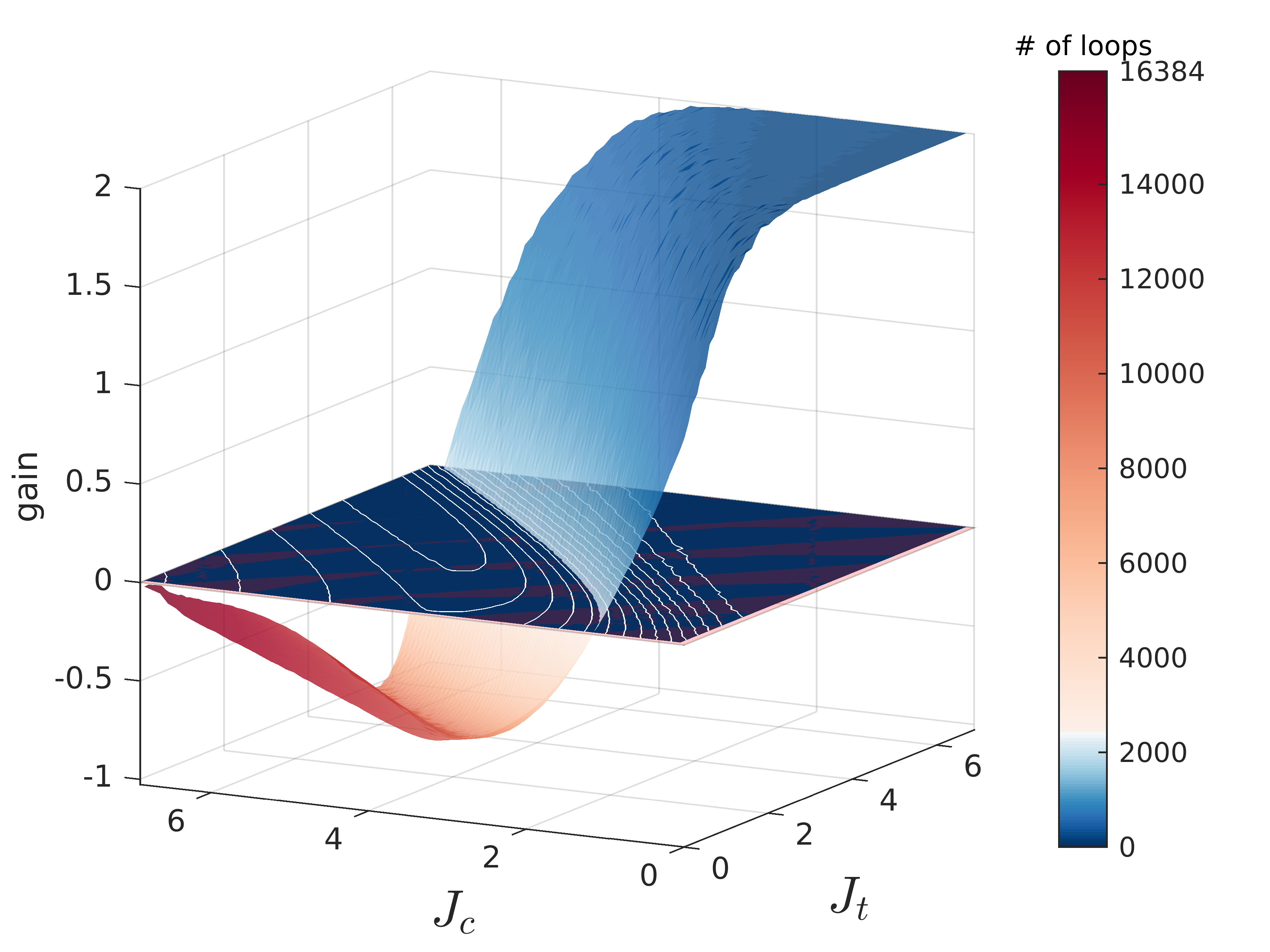}
\caption{Gain $g=2(n-n_\mathrm{eq})/(n+n_\mathrm{eq})$, where $n$ is the steady-state number of loops in presence of a splitter and $N_\mathrm{eq}$ the number of loops in equilibrium. Color code gives $n$. White lines indicate gains of -1.0, -0.8,  $\ldots$, 1.8. System size is $N=256$. Note that the gain is maximal at low loop numbers. }\label {fig:comparison}
\end{figure}

We calculate the gain $g=2(n-n_\mathrm{eq})/(n+n_\mathrm{eq})$, Fig.~\ref{fig:comparison}, where $n_\mathrm{eq}$ is the number of loops in equilibrium. Clearly, $g$ varies from $-2$, when $n/n_\mathrm{eq}=0$, through 0, when $n/n_\mathrm{eq}=1$, to 2, when $n_\mathrm{eq}/n=0$. Since loops contain exclusively cis-joins, we find that $g \to 0$ in the limit $J_c \gg J_t$, where the formation of trans-joins is suppressed and splitters encounter no chains. In the opposite limit of $J_t \gg J_c$, the number of chains at equilibrium reaches a maximum. Here, the ability of splitters to disassemble these `mistakes' and `recycle' them into the MC algorithm drives the gain to saturate at its maximal value, $g = 2$. Between these two limits, $g(J_c, J_t)$ is non-monotonic, displaying a minimum of $g \approx -1$, corresponding to $n_\mathrm{eq}=3n$. 

The non-monotonicity can be understood by considering the limit $J_c^{-1} = J_t = 0$, where the equilibrium state and non-equilibrium steady state are both pure loops, and $g =0$. Decreasing $J_c$ to a finite but still very large value will give an occasional `broken loop' with a single trans join, Fig.~\ref{fig:model}(e). The `defective' join has near unity probability of be `corrected' to reform a loop if chosen in a MC step, but a splitter acting on a `broken loop' will only re-form a loop with probability $(1/4)^4 \approx 0.004$. We therefore expect $g$ to become negative.

Focus now on the locus  $J_{c}^{*}(J_t)$ where the surface in Fig.\ref{fig:comparison} intersects the $g = 0$ plane. When $J_c<J_{c}^{*}$, active self-disassembly enhances the yield of loops ($g>0$). The function $J_{c}^{*}(J_t)$ is non-linear: splitters do not simply rescale the temperature and change $J_c$ and $J_t$ by the same factor. We note that there is a region at small $J_t$ where $g>0$ even when $J_c > J_t$: in other words, active self-disassembly can be beneficial even when thermodynamics already favours cis joins. Quantitatively, $J_c < J_{c}^{*}(J_t)$ for $J_t\lesssim 3.4$. 

To understand why splitters enhance loop formation beyond the thermodynamic advantage conferred by $J_c > J_t$ in the limit of small $J_c$ and $J_t$, consider our system at $J_c = J_t = 0$. At thermal equilibrium, there is a small but non-vanishing concentration of loops coexisting with monomeric angles and various `mistakes' (predominantly short chains). Splitters do not act on the loops, but can disassemble chains to enhance loop formation, either by reforming a loop immediately, or via further MC steps. So, we may expect $g>0$ at the origin and, by extension, in its vicinity, as observed. An alternative, more detailed, understanding of the enhancement of loops by splitters in this region of parameter space can be obtained by defining `precursor' states that are one step away from loops~\cite{SM}.

In summary, we have introduced a lattice model to study the effect of non-equilibrium disassembly on the yield of equilibrium self-assembly in forming a target structure: 4-loops with only cis-joins, Fig.~\ref{fig:model}. One might have expected that active self-disassembly enhances the yield of such loops only if trans-joins are thermodynamically favoured, \textit{i.e.}, when $J_t > J_c$; in other words, when there is an abundance of chains for splitters to disassemble chains and generate substrates for assembling more loops. Instead, we find that when $J_t \lesssim 3.4$, active self disassembly enhances the yield of loops even when $J_c > J_t$. In particular, at $J_c = J_t = 0$, $g \approx 0.73$, corresponding to $n/n_{\rm eq} \approx 2.15$, \textit{i.e.}, a 115\% enhancement of loops. 

This feature is robust against increasing the number of splitters, or equialently, increasing the frequency of splitter action per MC step. Doing so somewhat increases $g$ but does not change the region of parameter space in which $g>0$~\cite{SM}. Specifically, within our uncertainties, the value of $J_c$ giving $g=0$ is independent of the frequency of splitter action per MC step. Furthermore, the features reported above are essentially independent of system size $N$ as an analysis for $N=2$ shows~\cite{SM}.

The motivation for our study is the ubiquitous  coupling of active self-disassembly to self-assembly in biology, but our model is not directly bio-mimetic. Nevertheless, it is fruitful to place our findings in a biological context. Our splitters actively disassemble chains and thereby enhance the yield of loops above that expected from the Boltzmann factor alone. This is a form of proofreading which, like that first suggested by Hopfield~\cite{Hopfield1974} and Ninio~\cite{Ninio1975}, involves {\it driven} self-disassembly that breaks detailed balance, which in any physical implementation requires energy dissipation. So, both are instances of `active proofreading'~\footnote{`Kinetic' can mean `active, dynamic, full of energy' (Oxford English Dictionary); one author names Hopfield and Ninio's process `active proofreading'~\cite{Zuckerman}.}. 

However, our scheme is conceptually distinct from Hopfield-Ninio's~\cite{Hopfield1974,Ninio1975}, where active self-disassembly {\it follows} a quasi-equilibrium step of self-assembly, adding a (crucial) delay into the process to enhance binding specificity~\cite{Alon2019}. In our case, active self-disassembly occurs {\it in parallel} with equilibrium self-assembly, with the former feeding back building blocks for the latter, enhancing the yield of target structures. 

A closer analogue is `actin treadmilling'~\cite{Wegner:1976ks,Carlier2017}. An actin filament forms via the simultaneous near-equilibrium self-assembly of ATP-bound actin monomers at the `barbed end' and the self-disassembly of such monomers from the `pointed end' driven by non-equilibrium ATP hydrolysis, which recycles monomers for self-assembly. As in our scheme, simultaneous active self-disassembly and equilibrium self-assembly modifies the system's statistics, giving a peaked filament length distribution~\cite{Erlenkamper:2013cy,Kondev2016}. By contrast, in equilibrium, one has either indefinite growth or an exponential length distribution~\cite{Kondev2016}. The conceptual similarities with our model are evident, as obtaining a target filament length is akin to obtaining loops in our model. Significantly, an evolutionary process selecting kinetic parameters to reach a given target length was found to yield active self-disassembly~\cite{Hadjivasiliou2023}.

We have not optimised the action of our splitters in any way. So, while they do not make mistakes in the sense that they never split loops, they do not disassemble all unwanted structures, because they recognise chains but not non-square loops, although such loops hardly ever appear. Separately, by associating an energy cost with splitter action, perhaps proportional to the number of joins to be broken, \textit{i.e.}, the chain length, we can investigate how to minimise dissipation and how this is coupled to maximising $g$. Our model does not map to real time. To explore the optimisation of the speed to reach steady state, we need to take realistic account of the energy barriers to reconfigure at equilibrium and include an increase in disassembly time with chain length when a splitter acts. Implementing such optimisation using a genetic algorithm may throw light on how biological self-disassembly has evolved~\cite{Hadjivasiliou2023}. 

To conclude, we return to the finding that active self-disassembly is able to enhance the yield of loops above that achievable by equilibrium self-assembly alone in the limit $J_c, J_t \to 0$.  This is interesting, because the importance of weak, non-covalent interactions such as hydrogen bonds and the van der Waals force in biology has long been recognised~\cite{Frieden1975,Vladilo2018}: strong binding limits flexibility and reversibility, and high selectivity can in fact be better achieved using multiple weak bonds rather than a single strong bond~\cite{Knapp2008,Frenkel2011,Prins2020}. It is therefore interesting that active self-disassembly in our model allows an enhancement of a target self-assembled structure {\it without} having to resort to significant increase in the strength of the `correct' bonds (cis-joins). Future work will show whether this feature is in fact generic.

\begin{acknowledgments}
We thank Jacques Rougement and Patrick Warren for very helpful discussions. KK and WCKP would like to thank the Isaac Newton Institute for Mathematical Sciences, Cambridge, for support and hospitality during the program `New statistical physics in living matter', where work on this paper started. This work was supported by EPSRC grant no EP/R014604/1. KK was partially supported by SNSF grants CRSII5-183550 and 200020E-219164. JPE was partially supported by NCCR SwissMAP.
\end{acknowledgments}

For the purpose of open access, the authors have applied a Creative Commons Attribution (CC BY) licence to any Author Accepted Manuscript version arising from this submission.

\providecommand{\noopsort}[1]{}\providecommand{\singleletter}[1]{#1}%

\end{document}